%% file: CEPless-main-IEEE.tex
\documentclass[conference]{IEEEtran}
\IEEEoverridecommandlockouts
\IEEEpubid{\makebox[\columnwidth]{978-1-7281-6251-5/20/\$31.00~\copyright2020 IEEE \hfill} \hspace{\columnsep}\makebox[\columnwidth]{ }}
\input{sections/common}

\newboolean{komversion}
\setboolean{komversion}{true} 
\ifthenelse{\boolean{komversion}}{
    \usepackage{eso-pic}
    \AddToShipoutPictureBG{
        \AtPageUpperLeft{%
            \raisebox{-3\baselineskip}{
\makebox[\paperwidth]{\colorbox{yellow!40}{\begin{minipage}{0.85\paperwidth}\centering\small
                            {Manisha Luthra, Sebastian Hennig, Kamran Razavi, Lin Wang, Boris Koldehofe: \emph{Operator as a Service: Stateful Serverless Complex Event Processing}. In the Proceedings of IEEE International Conference on Big Data, December, 2020}
                        \end{minipage}}}}%
                    }
                    \AtPageLowerLeft{%
                        \raisebox{3\baselineskip}{
\makebox[\paperwidth]{\fbox{\begin{minipage}{0.85\paperwidth}\scriptsize
                                        {The documents distributed by this server have been provided by the contributing authors as a means to ensure timely dissemination of scholarly and technical work on a non-commercial basis. Copyright and all rights therein are maintained by the authors or by other copyright holders, not withstanding that they have offered their works here electronically. It is understood that all persons copying this information will adhere to the terms and constraints invoked by each author's copyright. These works may not be reposted without the explicit permission of the copyright holder.}
                                    \end{minipage}}}}%
                    }
    }
}{}

\begin{document}
\title{Operator as a Service: Stateful Serverless Complex Event Processing}

\author{\IEEEauthorblockN{Manisha Luthra\IEEEauthorrefmark{1},
Sebastian Hennig\IEEEauthorrefmark{1}, Kamran Razavi\IEEEauthorrefmark{1}
}
\IEEEauthorblockA{
\IEEEauthorrefmark{1}\textit{Technical University of Darmstadt, Germany} \\
\{firstname.lastname\}@kom.tu-darmstadt.de \\
razavi@tk.tu-darmstadt.de
}
\and
\IEEEauthorblockN{Lin Wang\IEEEauthorrefmark{1}\IEEEauthorrefmark{2},
Boris Koldehofe\IEEEauthorrefmark{1}\IEEEauthorrefmark{3}}
\IEEEauthorblockA{
\IEEEauthorrefmark{2}\textit{Vrije Universiteit Amsterdam, Netherlands} \\
lin.wang@vu.nl
}
\IEEEauthorblockA{
\IEEEauthorrefmark{3}\emph{University of Groningen, Netherlands} \\
b.koldehofe@rug.nl
}
}

\maketitle

\input{sections/abstract}


\input{sections/introduction}
\input{sections/case-study}

\input{sections/overview}
\input{sections/model}

\input{sections/design}

\input{sections/implementation}

\input{sections/evaluation}
\input{sections/related-work}
\input{sections/conclusion}
\balance
\bibliographystyle{abbrv}
\bibliography{cep-less}

\end{document}

%% file: sections/common.tex
\usepackage{booktabs} 
\usepackage{comment}
\usepackage{url}
\usepackage{amsfonts}
\usepackage{pifont}
\usepackage{caption, setspace}
\usepackage{balance}
\usepackage[linesnumbered,ruled,vlined,lined,boxed]{algorithm2e}
\usepackage{algorithmic}
\usepackage[acronym,toc,shortcuts]{glossaries}
\usepackage{amsthm}
\usepackage{color}
\usepackage{soul}
\usepackage{footmisc} 
\usepackage{todonotes}

\definecolor{teagreen}{rgb}{0.82, 0.94, 0.75}
\sethlcolor{teagreen}

\newcommand{\codebox}[1]{\texttt{\colorbox{teagreen}{#1}}}
\theoremstyle{definition}
\newtheorem{definition}{Definition}[section]

\theoremstyle{remark}

\usepackage[]{todonotes} 
\makeatletter
\newcommand\footnoteref[1]{\protected@xdef\@thefnmark{\ref{#1}}\@footnotemark}
\makeatother

\usepackage{cleveref}

\SetKwProg{Function}{function}{}{end function}
\SetKwInOut{KwVar}{Variables}
\SetKwInOut{KwDescribe}{Description}
\SetKw{KwTo}{to}
\SetKw{KwAnd}{and}
\SetKwIF{If}{ElseIf}{Else}{if}{then}{else if}{else}{end if}
\SetKwFor{For}{for}{do}{end for}
\SetKwFor{ForEach}{for each}{do}{end for}
\SetKwFor{ForAll}{for all}{do}{end for}
\SetKwFor{ForAllP}{for all}{do in parallel}{end for}

\usepackage{color}
\usepackage{float}
\usepackage{subfig} 
\usepackage{listings} 
\newcommand{\system}{\textsc{CEP\-less}\xspace}
\newcommand{\eg}{e.g.,\xspace}

\newcommand*\circled[1]{\tikz[baseline=(char.base)]{
            \node[shape=circle,draw,inner sep=0.3pt,color=white,fill=black] (char)
            {#1};}}

\lstdefinelanguage{Scala}{
	morekeywords={},
	morekeywords=[2]{
		WHERE,JOIN,ON,WINDOW,SLIDING,NOT,
		DEMAND,LATENCY,DEMAND,PROXIMITY,WHEN,
		abstract,case,catch,class,def,
		do,else,extends,false,final,finally,
		for,if,implicit,import,lazy,match,mixin,
		new,null,object,override,package,
		private,protected,requires,return,sealed,
		super,this,throw,trait,true,try,
		type,val,var,while,with,yield,
		window, join, on, where, within, demand, latency, proximity, when, Vector, def, val, for},
	sensitive=true,
	morecomment=[l]{//},
	morecomment=[n]{/*}{*/},
	morecomment=[s][identifierstyle]{`}{`},
	morestring=[b]",
	morestring=[b]',
	morestring=[b]"""
}

\lstdefinelanguage{JavaScript}{
	morekeywords={
		break,const,continue,delete,do,while,export,for,in,function,
		if,else,import,in,instanceOf,label,let,new,return,switch,this,
		throw,try,catch,typeof,var,void,with,yield
	},
	sensitive=true,
	morecomment=[l]{//},
	morecomment=[s]{/*}{*/},
	morestring=[b]",
	morestring=[d]'
}

\lstdefinelanguage{docker}{
    morekeywords={image, environment, depends_on, ports, volumes, networks, deploy, replicas, restart_policy,
    condition, placement, constraints, privileged},
    sensitive=true
}

\lstdefinestyle{interfaces}{
  float=tp,
  floatplacement=tbp,
  abovecaptionskip=-5pt
}

\lstnewenvironment{codenv}{\lstset{language=Scala, docker}}{}
\lstnewenvironment{framedcodenv}{\lstset{language=Scala,frame=btlr}}{}
\makeglossaries
\newacronym{CEP}{CEP}{Complex Event Processing}
\newacronym{TCEP}{\textsc{Tcep}}{Transitive-CEP}
\newacronym{OP}{OP}{Operator Placement}
\newacronym{QoS}{QoS}{Quality of Service}
\newacronym{DCEP}{DCEP}{Distributed Complex Event Processing}
\newacronym{GA}{GA}{Genetic Algorithms}
\newacronym{IoT}{IoT}{Internet of Things}
\newacronym{MFGS}{MFGS}{Moving Fine-Grained State}
\newacronym{SMS}{SMS}{Seamless Minimal State}

%% file: sections/abstract.tex

\begin{abstract}
Complex Event Processing (CEP) is a powerful paradigm for scalable data management that is employed in many real-world scenarios such as detecting credit card fraud in banks.
The so-called \emph{complex events} are expressed using a \emph{specification language} that is typically implemented and executed on a specific \emph{runtime system}.
While the tight coupling of these two components has been regarded as the key for supporting CEP at high performance, such dependencies pose several inherent challenges as follows.
(1)~Application development atop a CEP system requires extensive knowledge of how the runtime system operates, which is typically highly complex in nature. (2)~The specification language dependence requires the need of domain experts and further restricts and steepens the learning curve for application developers.

In this paper, we  propose \system,
a scalable data management system that decouples the specification from the runtime system by building on the principles of \emph{serverless computing}.
\system provides ``operator as a service'' and offers flexibility by enabling the development of CEP application in \emph{any} specification language while abstracting away the complexity of the CEP runtime system.
As part of \system, we designed and evaluated novel mechanisms for \emph{in-memory processing} and \emph{batching} that enable the
stateful processing of CEP operators even under high rates of ingested events.
Our evaluation demonstrates that \system can be easily integrated into existing CEP systems like Apache Flink while attaining similar throughput under high scale of events (up to 100K events per second) and dynamic operator update in \textasciitilde238~ms.
\end{abstract}

%% file: sections/introduction.tex

\section{Introduction} \label{sec:Introduction}
Complex event processing (CEP) is a data management paradigm used in a wide range of applications to efficiently detect interesting event patterns in event streams.
Such event patterns often named \emph{complex events}, allow applications upon their detection to adapt to situational changes, such as the detection of fraud in credit card payments~\cite{apama} and deriving tweet trends in Twitter~\cite{apacheHeron}.
The strengths of CEP reside in the simple specification of complex events by means of a query language and the support for efficient and distributed execution of the event detection logic.

Therefore, almost every CEP system provides two key components: (i)~the \emph{specification language} used to define event patterns and (ii)~\emph{the runtime system} to execute the event detection logic.
Typically, these two components are highly intertwined.
The constructs that describe the event detection logic are mapped to specific, at times infrastructure dependent, operator implementations, e.g., the algorithms for detecting sequences of events in a time-based window.
Driven by preferences of programmers and the underlying systems infrastructure, many distinct CEP systems have been proposed~\cite{apacheHeron, flinkCarbone, akidau2013millwheel, beam}, offering each very specific features having specific programming models and infrastructures in mind. For example, classic CEP programming models such as CQL~\cite{cql/vlbd/arasu2006} and SASE~\cite{sase/sigmod/eugene2006} are based on SQL-like semantics, and hence, they also share many limitations of SQL. 
In particular, it is very difficult to express complex business logic using these programming models.
Rather in current practices such as Google Dataflow~\cite{googleDataflow}, Millwheel~\cite{akidau2013millwheel}, and Flink~\cite{flinkCarbone}, object-oriented languages are often used to express complex business logic in the form of user-defined functions (UDFs).
With UDFs an operator can encapsulate any business logic and hence can be customized as per user needs.

While the expressiveness is significantly improved with UDFs, existing CEP system still fall short in several aspects.
One of them is the lack of runtime independence.
Although within each CEP system, queries can be specified in a highly composable manner or even altered, there is little support to benefit from reusing development effort from one CEP-application to another.
Simply, rewriting the query specification from one system to another is difficult since the way CEP queries are written has specific execution semantics in mind which are known to diverge between multiple systems~\cite{Cugola2012}.
Furthermore, the support for dynamic updates to the UDFs is weak.
Thus, changes to the implementation of specific operators or the definition of new functionality often require a restart and new deployment of the operators~\cite{flinkCarbone}, which is problematic in many applications like fraud detection where system availability is critically important.

In this paper, we aim for a better understanding of how one can benefit from the diversity of different CEP systems
and enhance their applicability to a wide spectrum of different infrastructures.
By proposing a data management system that builds on the principles of serverless computing architecture~\cite{hendrickson2016serverless}, we aim to enhance the reuse and integration of CEP operators.
Furthermore, by proposing methods for adding new functionalities and altering the operator logic at run-time, CEP systems can adapt their processing logic dependent on the application context as well as the features of the underlying infrastructure.
This way, the diversity of operator implementations becomes no longer an obstacle in the development cycle, but a feature that allows CEP systems to evolve to new requirements and adapt to contextual changes at run-time.

While existing serverless computing platforms~\cite{lambda, googleCloudFunctions} provide important concepts for the scalable execution of operator implementations, the extension of CEP systems to a serverless platform imposes many challenges.
First, CEP operators often perform stateful processing such as detection of correlated events within a time-based window of the data stream~\cite{Cugola2012}.
Stateful processing is not easily possible in a serverless model because conventionally each function or operator execution is required to be isolated and ephemeral in current platforms. For example in AWS Lambda, the functions have a limited lifetime of up to 15 minutes, which is the maximum among several serverless platforms. For the functions or operators, it is assumed that the state is not recoverable across invocations.
This stays in contradiction to the lifetime of a CEP operator which is required to be running as long as the query is executed~\cite{Hellerstein2019}.
Second, the execution mechanisms in current CEP systems perform many optimizations such as flow control, backpressure, and in-memory network buffers to guarantee low latency and high throughput.
Achieving equal performance in an existing serverless platform is currently not possible especially due to missing optimizations as aforementioned and slow communication through storage.
For example, in AWS Lambda~\cite{lambda} it is only possible to communicate between two lambdas using S3 that is extremely slow and more expensive than point-to-point networking~\cite{serverless/socc/eric2017}.

\emph{Contributions:}
To overcome these flexibility limitations of current systems, we propose a new CEP system named \system based on the \emph{function as a service (FaaS)} concept of \emph{serverless computing}~\cite{lambda}.
In this system, we provide CEP \emph{operators as a service} which means that the operator thus can be specified independently of the underlying {runtime system} without any side-effects to the other system dependent operators running in the CEP system.
The approach enables \emph{specification language independence} from the CEP runtime system for the specified operators, which essentially means that the operators can be developed in \emph{any} programming language desired by the developer.
Furthermore, we contribute to \system with two mechanisms
(i)~supporting the stateful processing of events, which poses a challenge in existing serverless platforms and (ii)~meeting the latency and throughput requirements of the CEP applications.
\system addresses these challenges through \emph{in-memory queue management}, which maintains the state in an in-memory queue for stateful operators like time-based windows. Furthermore, the \emph{batching mechanism} aids in providing high throughput and low latency in event processing.
Overall, \system introduces a high degree of flexibility for CEP systems with the involvement of FaaS in conjunction with our advanced design.
In summary, this paper makes the following contributions:
\newline

\begin{enumerate}
    \item We propose mechanisms for \emph{in-memory queue management} and \emph{batching} that enable stateful processing and ensure correctness and fast delivery of events which are extremely important for CEP systems.
    \item We introduce a unified \emph{user-defined operator interface} that allows the integration of highly diverse CEP runtime systems into \system system and allows runtime system independence.
    \item We implement and evaluate \system on two state-of-the-art CEP systems \emph{Apache Flink}~\cite{flinkCarbone} and \emph{TCEP}~\cite{Luthra2018} using an open and anonymous credit card transaction dataset~\cite{transactionsDataset}.
    Results show that \system enables runtime system independent updates of user-defined operators while attaining equal throughput and preserving low latency overhead (\textasciitilde1.9~ms) under high scale of event rates of up to 100K events per second.
\end{enumerate}

The rest of the paper is structured as follows.
\Cref{subsec:example} presents a motivational example.
\Cref{sec:model} presents the system model introducing the important system entities. \Cref{sec:design} describes \system system design.
\Cref{sec:evaluation} shows the evaluation.
\Cref{sec:relatedwork} discusses the related approaches and \Cref{sec:conclusion} concludes the paper.

%% file: sections/case-study.tex

 \begin{figure*}
\centering
\includegraphics[width=0.8\linewidth]{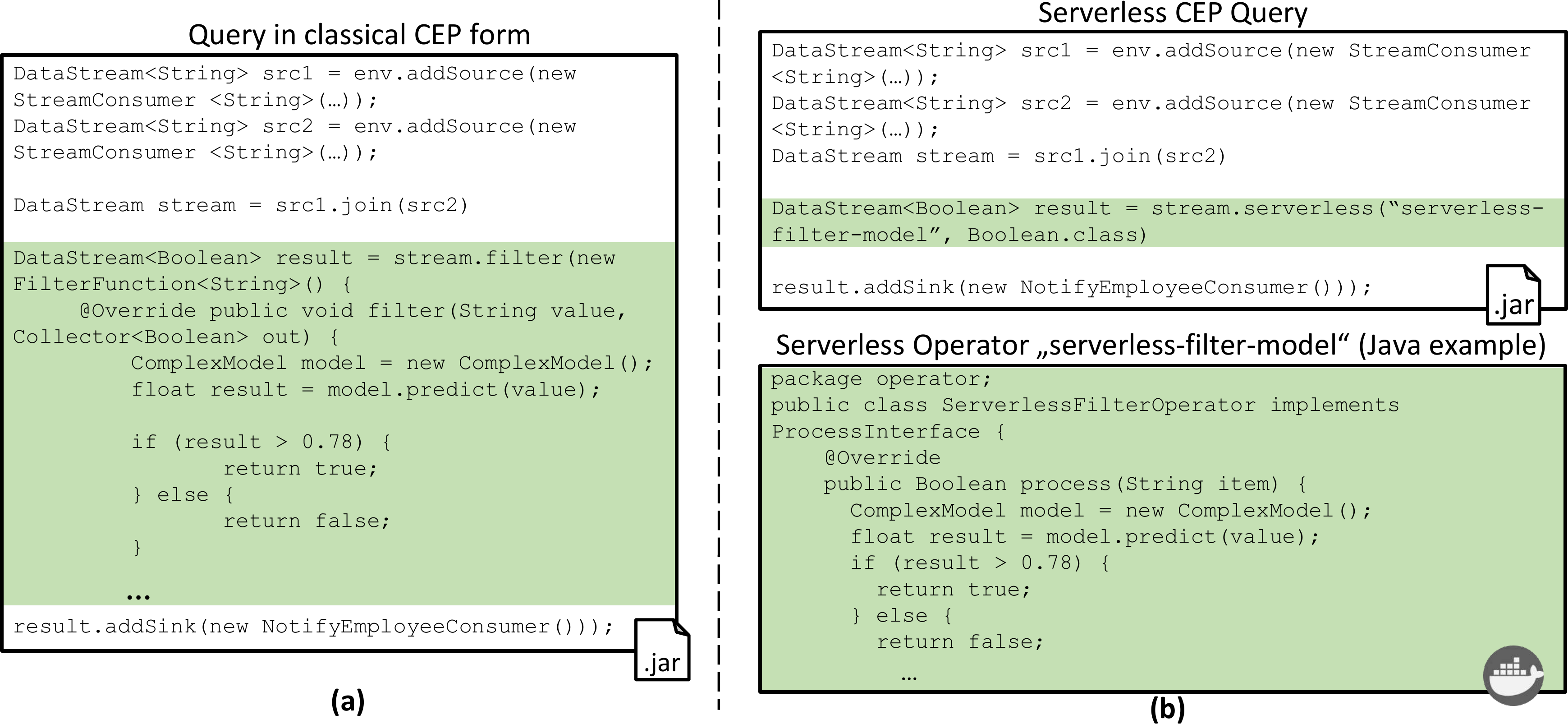}
\setlength{\abovecaptionskip}{5pt}
\setlength{\belowcaptionskip}{-15pt}
\caption{An example of a simple  \codebox{filter} application for fraud detection (a) in traditional CEP systems and (b) in the \system system. Here,  the \codebox{serverless-filter-model} in (b) is decoupled from the CEP runtime and can be updated on the fly.}
\label{fig:case-study-intro}
\end{figure*}

\section{Problem Statement Using Fraud Detection Example}~\label{subsec:example}

 A financial institution wants to detect payment frauds in the real-time credit card transactions of its customers.
 The fraud detection algorithm includes complex business logic such as machine learning models.
 Typically, this algorithm is required to be dynamically updated to incorporate newly observed transaction patterns of fraud.
 The department has its proprietary machine learning library that is implemented in a highly efficient and scalable language such as Rust, or a proprietary language developed within the department.
 The fraud department uses a CEP system for detecting fraud in sub-second latency and for a high number of users, resulting in a high scale of incoming transactions to be processed.
 Current CEP systems, however, fail to fulfill the above requirements of expressing a complex business logic as an operator without any prior knowledge of the CEP runtime system, while using the language preference of the department for the specification of complex events.

 To better understand the requirements, consider an example specification of fraud detection in a widely used CEP system Apache Flink~\cite{flinkCarbone} (cf. \Cref{fig:case-study-intro}(a)).
 Here, the \codebox{highlighted} code refers to the specification of the operator or business logic that uses a pre-trained machine learning model to detect fraud\footnote{In the example, we have kept the business logic rather simple for better understandability.}.
 We can see that the specification is tightly coupled to the runtime APIs like \texttt{DataStream} and \texttt{FilterFunction}. Furthermore, the query operator \codebox{filter} is ossified and cannot be updated after deployment.
 Besides, the language of preference cannot be used due to missing APIs.
 This dependency on a specific runtime system can be problematic because of several reasons:
(1) CEP systems are extremely complex by nature and
the respective department for the development of fraud detection operators in a financial institution may not have the expertise to deal with such complex systems.
(2) Fraud detection might require extending the existing runtime with external modules, e.g., machine learning libraries which can be very cumbersome in current CEP systems, if not impossible.
(3) Different financial institutions might have individual specification language preferences for fraud detection operators. This is also not easily possible without extending the complete CEP system for a different specification language or rewrite of the operators.

 On the right side \Cref{fig:case-study-intro} (b) is our proposal, which segregates the business logic of the operator \codebox{serverless-filter-model}  from the CEP runtime by implementing the operator as a serverless function containerized in a virtualization environment. Hence, the independent \codebox{ServerlessFilterOperator} can be reused for different CEP runtime environment. This segregation leads to several research challenges that we address in this work.
 \begin{enumerate}
    \item How to allow dynamic updates of an operator, while guaranteeing high performance in the delivery of events?
     \item How to design user-defined operators, such as one for fraud detection, independent from the underlying CEP runtime, and the programming language?

 \end{enumerate}

%% file: sections/model.tex
\section{\system Model} \label{sec:model}
\system is able to consume continuous data streams $(D)$ from a set of event producers $(P)$ such as the Internet of Things devices. A set of event consumers $(C)$ express interest in inferring a complex event such as \emph{fraud detection} in the form of a query $q$. A query $q$ induces a directed acyclic \emph{operator graph} $G = (\Omega \cup P \cup C, D)$, where a vertex represents an operator $\omega \in \Omega$ and an edge represents the flow of events based on data streams, s.t., $D \subseteq (P \cup \Omega) \times (C \cup \Omega)$. Each operator $\omega$ dictates a processing logic $f_\omega$. 

\system provides a set of operators $\Omega = \{\Omega_S, \Omega_{UD} \}$ where an operator can be either a \emph{system-defined operator} $\omega_S \in \Omega_S$ or a \emph{user-defined operator} $\omega_{UD} \in \Omega_{UD}$. We define them as follows.

\begin{definition}{\emph{System-defined operators} ($\Omega_S$)}
is a set of standard CEP operators s.t. $\Omega_S \subseteq \Omega$ and $\omega_S \in \Omega_S$. Conventional CEP operators comprising of single-item such as \emph{selection}, logical operators such as  \emph{conjunction}, window operators such as \emph{sliding window}, and flow-management operators such as  \emph{join}~\cite{Cugola2012}.
\label{def:system_defined_operators}
\end{definition}

\begin{definition}{\emph{User-defined operators} ($\Omega_{UD}$)} is a set of user-defined operators s.t. $\Omega_{UD} \subseteq \Omega$ and $\omega_{UD} \in \Omega_{UD}$. Unlike system-defined operators, it contains custom logic or user-defined code that typically cannot be expressed by system-defined operators. The flow of information from $\omega_{UD}$ is encapsulated in a \emph{container}, which is detailed in the next section.
 \label{def:user_defined_operators}
 \end{definition}

In a distributed setting, the operators are typically placed on a set $N$ of nodes that are responsible to process the incoming data streams. The nodes hosting the $\omega_{UD}$ operator is managed by a \emph{node manager} $NM_{n}$ that handles all $\omega_{UD}$ operator incoming requests.

In \Cref{fig:systemmodel}, we illustrate an overview of this \system system model comprising of three layers: the serverless layer, the execution layer, and the devices layer. The devices layer is composed of primary devices such as the Internet of Things, which generates continuous data streams that are to be processed. The execution layer comprises of a variety of current CEP systems, which processes the data streams to derive a complex event. Lastly, the serverless layer provides system for defining user-defined operators and the flexibility of multiple runtime environments. The operator graph mapping to the layers in the figure comprises standard $\omega_S$ operators which are system-defined such as a stream operator ($\omega_{str}$) comprising continuous time-series event tuples in the form of $<E>$. Here, $E$ is a set of time-series event tuples $\{<ts_1, e_1>, <ts_2, e_2>, .. <ts_n, e_n>\}$.
In the \system system, multiple operator graphs can co-exist with different CEP systems in execution at the same time. For example, an operator graph of TCEP (with dotted lines) can co-exist with that of Flink (solid line) as represented in the figure.

\begin{table}[]
\small
\begin{tabular}{ll}
\hline
Notation & Meaning \\
\hline
  $P$       &  Set of event producers   \\
  $C$       &  Set of event consumers   \\
  $D$       &  Continuous data stream   \\
  $G$       &  Operator graph           \\
  $\Omega$  &   Set of CEP operators ($\omega \in \Omega$)  \\
  $\Omega_S$ &  Set of system-defined operators ($\omega_S \in \Omega_S$)  \\
  $\Omega_{UD}$ & Set of user-defined operators ($\omega_{UD} \in \Omega_{UD}$)  \\
  $N$       &  Set of nodes ($n \in N$) \\
  $NM_{n}$ & Node manager for each node $n$  \\
  $E$       & Set of events $(e \in E)$ \\
  $f_{\omega}$  &   Processing logic of an operator \\
  $C_{\omega}$  &   Operator container \\
  $eq_{\omega_{UD}}$ &  Event queue for each $\omega_{UD}$  \\
  $b_s$ and $b_r$   &   Sent and received batch of events  \\
  $t_{backoff}$ & Queue polling backoff time \\
\hline
\end{tabular}
\setlength{\abovecaptionskip}{0pt}
\setlength{\belowcaptionskip}{-5pt}
\caption{Notations and their meaning.}
\label{tab:notations}
\end{table}
\subsection{Container Model}
Operators containers provide the ability to encapsulate any processing logic $f_\omega$ to be used in CEP systems as a user-defined operator $\omega_{UD}$.

\begin{definition}{The \emph{operator container} ($C_{\omega}$)} serves as an interface for interacting with $\omega_{UD}$ in order to have a unified interaction between all operators and user applications. Events received by the CEP system in the execution layer are forwarded through the container interface into $\omega_{UD}$ which
provides an event entry point for $f_{\omega}$.
\label{def:container}
\end{definition}

While a user-defined operator continuously processes events, it can emit single or multiple events that are forwarded through the aforementioned interface back to the CEP system for further processing.

\subsection{Queue Model}
In order to provide a clean abstraction for $\omega_{UD}$ and enable flexibility of defining any processing logic $f_\omega$ as an operator, a messaging system needs to be established. To achieve this, we enable interactions between $\omega_{UD}$ and CEP systems using an \emph{event queue} as an event exchange point. We consider different aspects relevant for processing a continuous event stream such as (i) statefulness, (ii) performance, and (iii) in-order processing. There are no restrictions in place regarding the size of the queue as it is dependent on the number of event items processed.

\begin{figure}[]
    \centering
    \includegraphics[width=0.9\linewidth]{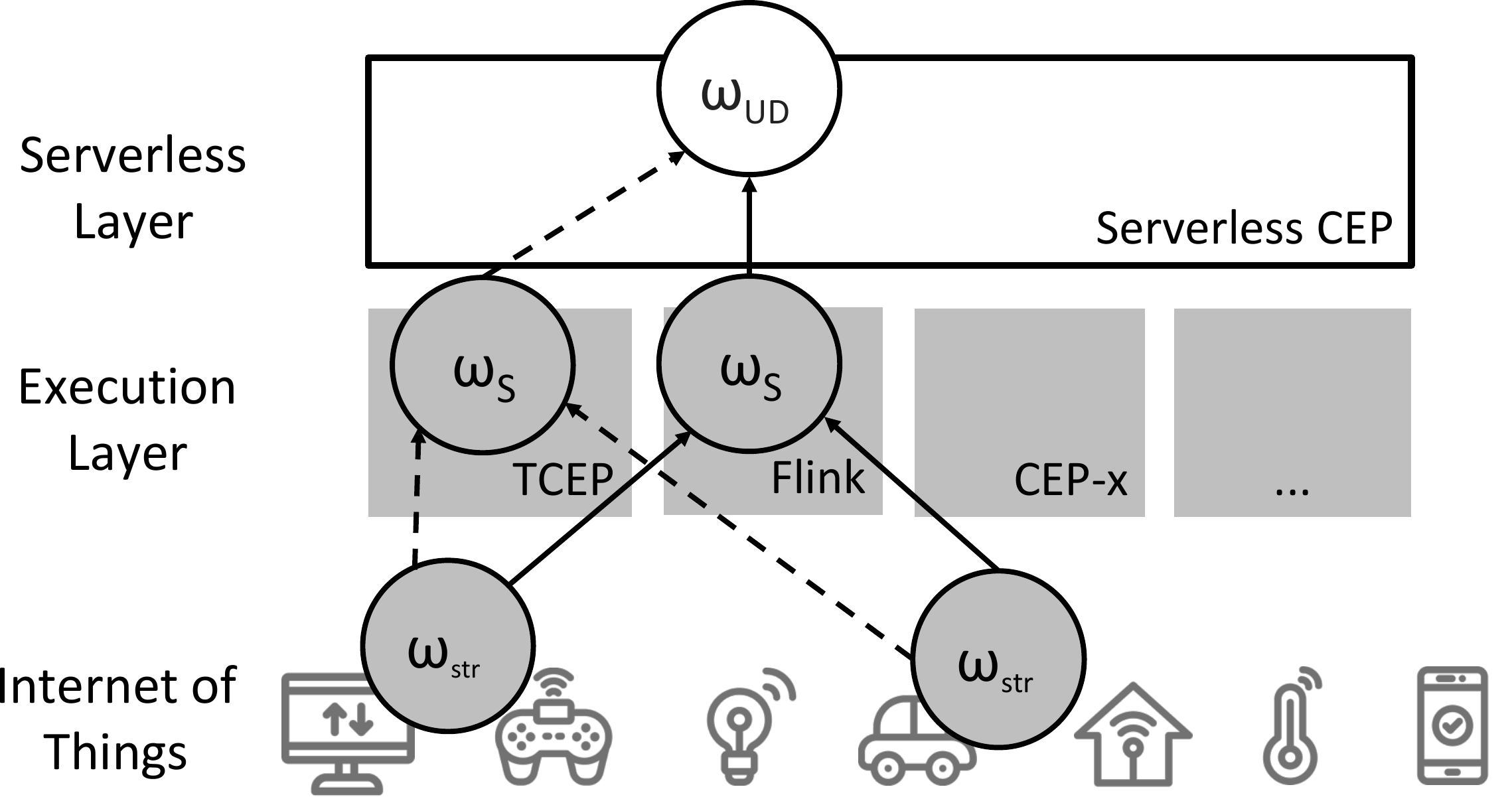}
    \setlength{\abovecaptionskip}{5pt}
    \setlength{\belowcaptionskip}{-5pt}
    \caption{The System model of \system.
    \textnormal{\textit{
    The grey layers are shared.}}}
    \label{fig:systemmodel}
\end{figure}

\begin{definition}{An \emph{event queue} ($eq_{\omega_{UD}}$)}\label{def:event-queue} is an instance in a queuing system distributed alongside with the CEP system on every node that participates in the cluster. Each $\omega_{UD}$ maintains two queues, namely \emph{Input} ($eq\_in_{\omega_{UD}}$) and \emph{Output} ($eq\_out_{\omega_{UD}}$). The former contains events received by the CEP system's operators $\omega_S$ which are to be forwarded for processing to $\omega_{UD}$. The latter contains the results of $\omega_{UD}$ execution, which are processed by the CEP system for the next operator in the operator graph. User-defined operators and CEP systems only communicate with the queue placed on their respective node.
Therefore, the queue can be accessed by user-defined operators and the CEP system through the local networking interface on the host which provides a negligible overhead in communication.
\label{def:queue}
\end{definition}

\begin{definition}{An \emph{event batch} ($b_s$ and $b_r$)}
comprises of a subset of event tuples from $E$ that are to be sent to ($b_s$) and from ($b_r$) the \system system through the event queues and the execution layer. 
\label{def:batch}
\end{definition}

\begin{definition}The {\emph{back-off interval} ($t_{backoff}$)}
is the time to backoff from polling the event queue $eq_{\omega_{UD}}$ when no events are received. It is incremented linearly at each polling step. This effectively reduces the execution overhead when the queues are empty.
\label{def:back_off}
\end{definition}

\begin{definition}{\emph{User-defined operator (UDO) interface}.} \label{def:udo-interface}
Queues interact with the system layer using a UDO interface. We consider this entity implemented in each CEP system enabling communication with $\omega_{UD}$ over event queues $eq_{\omega_{UD}}$. Each CEP system that wants to use \system is required to have an implementation of the UDO interface.
\label{def:udoi}
\end{definition}

 To support multiple $\omega_{UD}$ on a single machine, each $\omega_{UD}$ gets assigned two distinct event queues $eq_{\omega_{UD}}$ by \system as defined in \Cref{def:event-queue}. Incoming events in the first queue get processed by the $\omega_{UD}$  in a sequence of their arrival. The result of the processed events gets added to the second queue which is processed by the CEP system again.

%% file: sections/design.tex
\section{The \system system Design} \label{sec:design}
\begin{figure}[t]
    \centering
    \includegraphics[width=\linewidth]{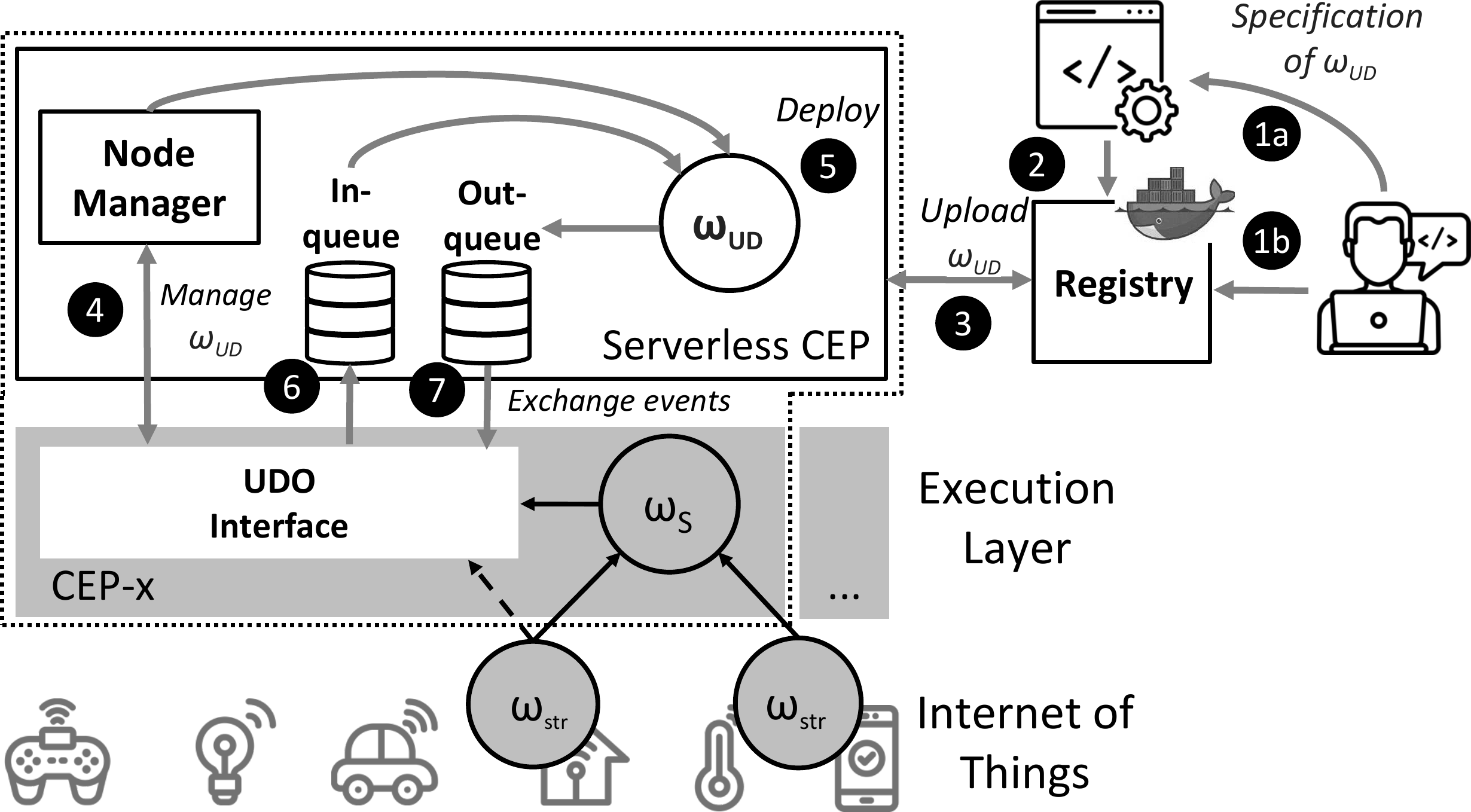}
    \setlength{\abovecaptionskip}{-2pt}
    \setlength{\belowcaptionskip}{-5pt}
    \caption{Overview of a node in the dotted line using \system. \textnormal{\emph{White layers are \system entities.}}}
    \label{fig:node-overview}
\end{figure}

\Cref{fig:node-overview} shows an overview of the \system design on a single node\footnote{On multiple nodes \system system is distributed (see dotted line).}. On the left side, we show the \system system components, which allow efficient integration of multiple CEP runtimes and their reuse while ensuring runtime updates of $\omega_{UD}$ operator.
We show developer-centric components on the right side, which allows developers to specify and submit $\omega_{UD}$ operators in their preferred programming language.
We show the flow of defining, deploying, and processing $\omega_{UD}$ using \system major components as follows.

We propose a programming interface to define and deploy specification language and CEP runtime independent user-defined operators $\omega_{UD}$ to the \system system.
It is important to note that at this point developers do not need to have any knowledge of the CEP runtime where the $\omega_{UD}$ operators will be deployed.
 \circled{1a} The developers can either program them using our interface or \circled{1b} reuse operators by packaging them into an executable operator container $C_\omega$ and submit to the \system registry.
\circled{\large{2}} The programming interface wraps the user-defined operator code into a container and pushes into the \system registry.
\circled{\large{3}} The \system system maintains all the $\omega_{UD}$ operators in a central registry to facilitate the reuse of operators.

To deploy a $\omega_{UD}$ operator, the \system system invokes the internal \emph{UDO interface} component of a specific CEP runtime. \circled{\large{4}} The UDO interface requests the deployment of the corresponding operator at a \emph{Node Manager} $NM_n$ instance, which is responsible for the deployment and registration of operator container $C_\omega$ on a node. \circled{\large{5}} After a $NM_n$ instance downloaded the requested $C_\omega$ from the  registry, it is started on the respective node for execution. The CEP runtime or system  communicates with the $\omega_{UD}$ over the \circled{\large{6}} input event queue $eq\_in_{\omega_{UD}}$ and \circled{\large{7}} output event queue $eq\_out_{\omega_{UD}}$ as soon as the deployment by $NM_n$ has finished. In this way, \system system provides a consistent functionality to other serverless systems~\cite{lambda, googleCloudFunctions} towards the user.
In the next subsections, we show the design of our system in terms of system components and their deployment. Furthermore, we introduce the interactions between \system and the provided execution layer, i.e., the CEP system.
In our design, we are focused on the aforementioned research questions identified before (cf.~\Cref{subsec:example}) in \Cref{sec:cepless-plattform} and \Cref{sec:in-memory-queue}, respectively.

\input{sections/serverless-platform}

%% file: sections/serverless-platform.tex
\subsection{Serverless Platform for CEP} \label{sec:cepless-plattform}
In consistent with the principles of serverless computing, the serverless platform of \system acts as a bridge between the user and the different CEP runtime environments. The platform simplifies the deployment and configuration for the user while still preserving the correct execution of the $\omega_{UD}$ operators on the underlying IoT resources. It manages every compute resource in the network and is therefore not bound to a specific system physical location, e.g., in co-location with a CEP runtime. Besides providing simple communication between the users and the CEP system, the platform also acts as a central point of knowledge in our deployment design. The platform keeps information of all the submitted $\omega_{UD}$ operators and operator execution requests by the CEP systems. In the following subsections, we detail on the sub-components used in the serverless platform of \system for deployment and runtime management of $\omega_{UD}$ operators.
\begin{lstlisting}[language=Scala, caption={User-Defined Operator (UDO) Interface.}
, label={listing:coi}]
trait UserDefinedOperatorInterface {
  def requestOperator(operatorName: String, cb: OperatorAddress => Unit): Unit
  def sendEvent(e: Event, address: OperatorAddress): Unit
  def addListener(address: OperatorAddress, cb: Any => Unit): Boolean
  def removeListener(address: OperatorAddress): Boolean
}
\end{lstlisting}
\subsubsection*{User-Defined Operator Interface} \label{sec:operator-interface}
We enable user-defined operators $\omega_{UD}$ to be utilized in CEP systems by abstracting the processing logic $f_{\omega}$ from the CEP system base. Operators can be implemented, used, and modified using any language that is executable by the virtualization layer. With this design choice, we give users the ability to use effectively any program that is executable in a virtualized container, with only minor modifications. For a CEP system to be able to communicate with this abstracted operator, it needs to implement an interface that handles events to and from the $\omega_{UD}$, equal to state-of-the-art serverless systems. Therefore, we propose the \textit{User-Defined Operator (UDO) Interface} which handles communication between the $\omega_{UD}$ and CEP system. Because every CEP system provides different language semantics, the implementation of this interface differs slightly between CEP systems, however, it is designed to provide only minimal overhead to the existing codebase.

Listing \ref{listing:coi} shows the required functionality that every UDO Interface needs in a CEP system to be usable with our extension. Line 2 is the first request issued by a CEP system that initiates the deployment of an operator by \system. Parameter \textit{operatorName} is a unique identifier for $\omega_{UD}$ to be deployed.
The second parameter is a function invoked as soon as $\omega_{UD}$ deployment was initiated and returns an operator address at which it is reachable. The operator address is used for routing events from the CEP system to $\omega_{UD}$ and has to be used at any interaction with the interface.
Line 3 shows the function that needs to be invoked to send an event to a user-defined operator. The parameter \emph{Event} is specific to the respective runtime system. The UDO interface is required to serialize the given event object into a readable format for the operator.
For the CEP system to be able to also receive resulting events from the deployed operator, it can add and remove a listener as seen in Line 4 and 5. This listener expects a function that is called every time a new event was received by \system from a $\omega_{UD}$ that has a listener registered.

\subsection{In-memory Queue Management} \label{sec:in-memory-queue}
The serverless platform provides in-memory queues $eq_{\omega_{UD}}$ for communication between $\omega_{UD}$ operators and the underlying execution layer comprising of different CEP runtimes.
The internal management of in-memory queue achieves three design goals (i) in order and stateful processing of events, (ii) runtime independent transfer of events to and from the platform and the execution layer and (iii) high performance in the delivery of events, which are extremely important for CEP applications.
In our design the CEP system and the \system are residing on the same computational resource, thus, have only minimal latency impact on the streaming system by avoiding network latency.
Of course, the design also allows us to deploy multiple user-defined operators as well in a distributed setting.

\begin{algorithm}[t]
\footnotesize
 \KwData{

 $b_s$ \qquad \qquad \quad $\leftarrow$ Batch of events sent (and processed) to $\omega_{UD}$\;
 $b_r$ \qquad \qquad \quad $\leftarrow$ Batch of events received from $\omega_{UD}$\;
 $t_{backoff}$ \qquad $\leftarrow$ Back-off interval increment\;
 $t_{sleep}$ \qquad \quad $\ \,\leftarrow$ Total current back-off time\;
 sendBuffer$ \quad \quad \leftarrow$ Internal Buffer for command compilation\;
 cmds $ \qquad \qquad \, \leftarrow$ Event Queue client commands\;
 inBatchSize\quad$\ \ \ \leftarrow$ Batch size of commands to be received\;
 outBatchSize $ \quad \leftarrow$ Batch size of commands to be sent\;
         }

\tcp{receives events from the execution layer ($\omega_S$)}
 \Function{receiveEvent} {  \label{line:startrcvEvent}
    $b_s$.push(event)\;  \label{line:endrcvEvent}
 }
\tcp{sends events to $\omega_{UD}$}
 \Function{backgroundThreadSend}  {  \label{line:startbackThread}
    $t_{sleep} \leftarrow 0$\;
     \While{true}{
      \eIf{$b_s$.size = 0}{   \label{line:emptyBatch}
        $t_{sleep} \leftarrow$ $t_{sleep} + t_{backoff}$ \; \label{line:increment}
        sleep($t_{sleep}$)\;
      }{
        \eIf{$b_s$.size $>$ outBatchSize}{  \label{line:exceedBatchSize}
            sendBuffer $\leftarrow b_s$.pop(0, outBatchSize)\;
        }{
            sendBuffer $\leftarrow b_s$\;
            $b_s$.clear()\;
        }
        \If{sendBuffer.size $>$ 0} {
            cmds $\leftarrow$ sendBuffer.compileCommands()\;
            cmds.flush()\; \label{line:flush}
        }
        $t_{sleep} \leftarrow 0$\;\label{line:endbackThread}
      }
     }
  }
\tcp{receives events from $\omega_{UD}$}
 \Function{backgroundThreadReceive}{  \label{line:startbackThreadRcv}
    $t_{sleep} \leftarrow 0$\;
     \While{true}{
      $b_r \leftarrow$ range(0, inBatchSize) \;\label{line:rangequery}
      \If{$b_r$.size = 0}{   \label{line:emptyBatchRcv}
        $t_{sleep} \leftarrow$ $t_{sleep} + t_{backoff}$ \; \label{line:incrementRcv}
        sleep($t_{sleep}$)\;
      }{
        forwardEvents($b_r$) \;
        $t_{sleep} \leftarrow 0$\;\label{line:endbackThreadRcv}
      }
     }
  }
 \caption{Event queue command batching and flushing}
  \setlength{\abovecaptionskip}{0pt}
 \setlength{\belowcaptionskip}{0pt}
 \label{alg:flush}
\end{algorithm}

\emph{In order and stateful processing of events.}
We use FIFO queuing mechanisms to ensure that the given data stream ingested to the CEP system in the execution layer keeps the order for processing of the operator. Equally, the data stream generated from $\omega_{UD}$ is handled in the same manner. In this way, \system keeps the order of events, however, \system is dependent on the execution layer ordering mechanisms, \eg Flink ensures ordering using watermarks, to ensure correctness in ordering such that there are no false positives and negatives.
We store the incoming events from the execution layer (to be sent to the $\omega_{UD}$ operator), outgoing events from the $\omega_{UD}$ operator (to be sent to the execution layer) as well as the intermediate state of operators in in-memory queues. These are maintained for every $\omega_{UD}$ operator by the serverless layer.

A universal message format encapsulates the events which are passed into the queue.
This format allows us to interact with different CEP runtimes without any dependence on their semantics.
As the events are received from the execution layer, the serverless layer communicates with the in-memory queue for sending and receiving the events.
The in-memory queue handles this communication using a client-server model.
When the client residing at the serverless layer creates a request to add a new event, the event is appended to the tail of a specified operator queue $eq_{\omega_{UD}}$. The communication is managed within the queue by issuing commands such as \texttt{push} and \texttt{pop} received in the form of TCP requests whenever the events are to be sent and received, respectively. The queues store the events and their state in memory without having to persist them on external storage, which could induce high transfer and I/O latency.
In this way, we provide fast but stateful processing of events, which is extremely important for CEP applications to deal with operators such as \emph{time-based windows}.

However, CEP systems often observe high rates of incoming events. For the client-server model as above, this would mean many TCP requests written to the network layer for each incoming event, which can be highly inefficient. We provide a solution to this problem as follows.

\emph{Guarantee high performance in the delivery of events.}
To handle the high rates of incoming events, we parallelize the process of event transfer and provide a batching mechanism that aids in processing a high rate of incoming events.
Both mechanisms provide advantages to process more events per time unit and therefore aid in provisioning high throughput in delivery. With batching we manage the command flushing mechanism to the TCP server, which is a known control mechanism of in-memory queues~\cite{redis}.  
This effectively means a suitable time when commands can be written to the network layer, i.e., the time when the commands are started. With automatic command flushing commands are always issued as soon as they are invoked at the client. This results in many commands (i.e., requests) being written to the network sequentially, resulting in many TCP requests opened and closed per time unit. To resolve this issue, we send commands by issuing command flushes at specified intervals, independently from the query process. With command flushs, the TCP connection is only opened once per batch, all the commands are sent and the result for \emph{all} commands is received. In \Cref{alg:flush}, we present this process. On the main thread (Line~\ref{line:endrcvEvent}) we collect the received events from the execution layer (or user-defined operator $\omega_{UD}$) and return immediately for fast processing in the execution layer. We utilize a background thread (Line \ref{line:startbackThread}--\ref{line:endbackThread}) immediately initialized in the beginning, which continuously processes the collected events by flushing the commands (Line \ref{line:flush}).
Instead of sending one request per event, we use a batching mechanism to collect multiple events and flush a batch $b_s$ of commands to the in-memory queue server after a certain threshold is exceeded (Line \ref{line:exceedBatchSize}). This essentially avoids the slowing down of the query as detailed above.

\begin{figure}
\centering
\includegraphics[width=0.38\linewidth]{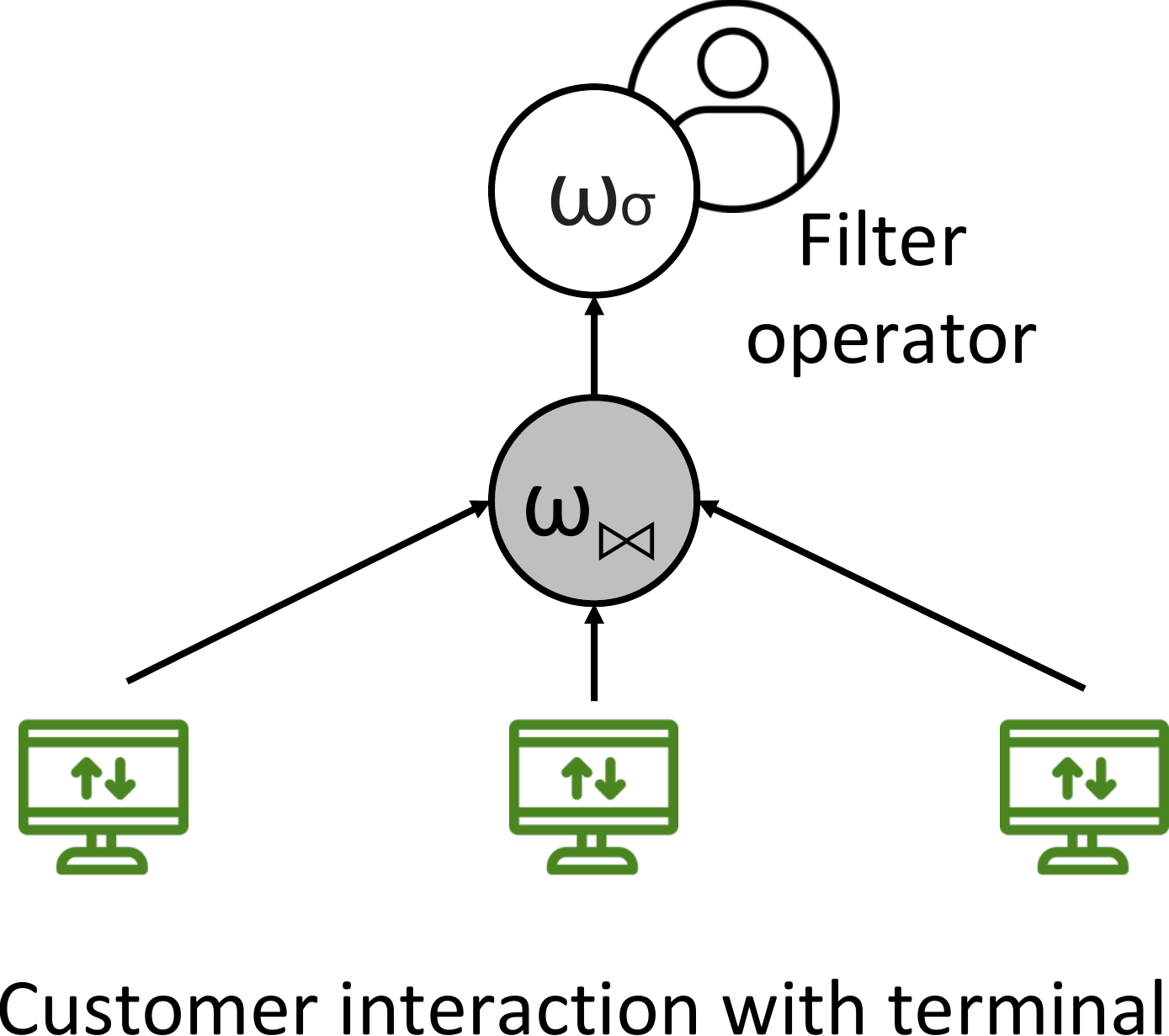}
\includegraphics[width=0.6\linewidth]{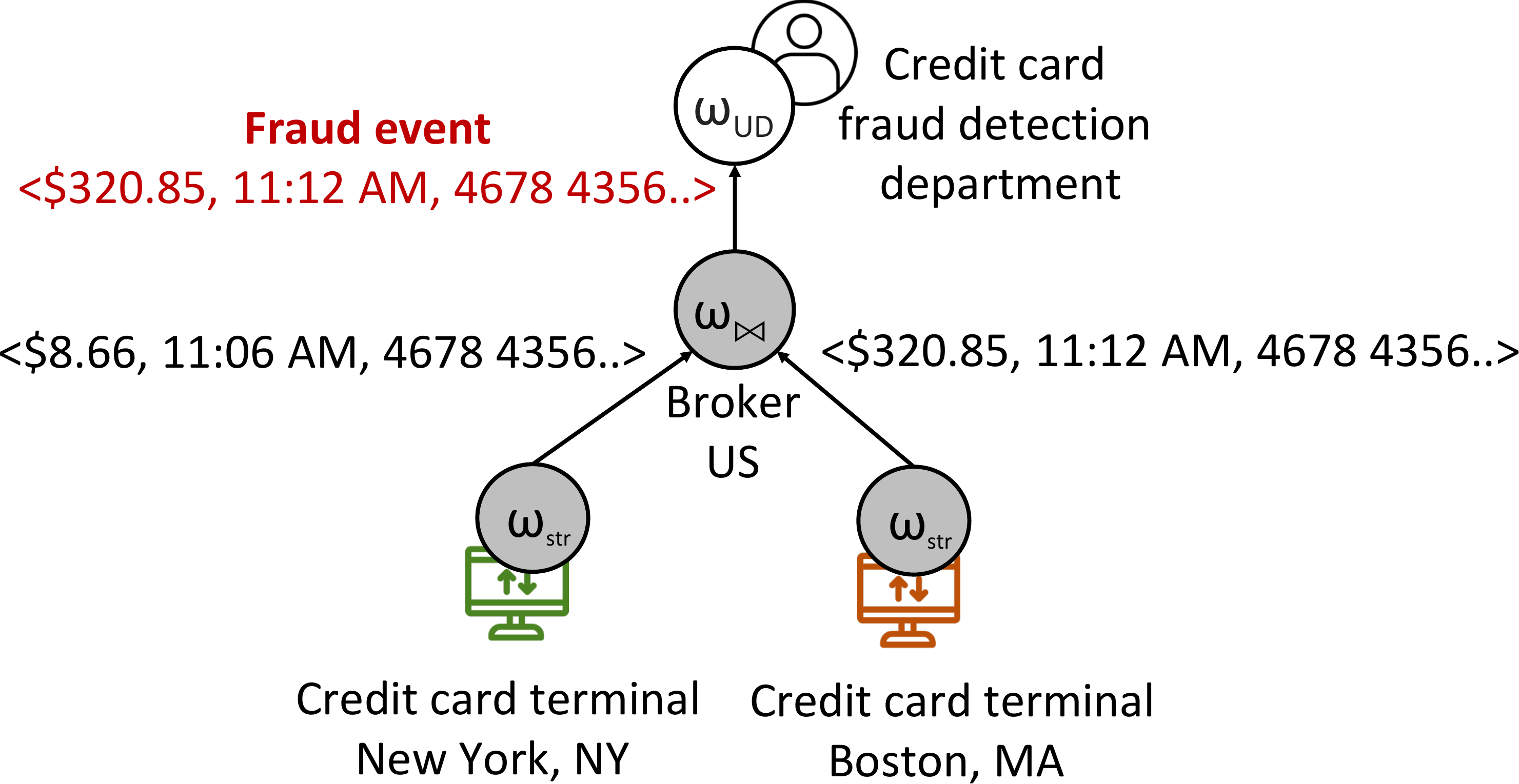}
\setlength{\abovecaptionskip}{0pt}
\setlength{\belowcaptionskip}{-5pt}
\caption{\label{benchmark-query}Credit card fraud detection query that is used for evaluating our system.}
\end{figure}
The background thread runs after a backoff time $t_{backoff}$ if previous batch $b_s$ was empty (Line \ref{line:emptyBatch}). The total backoff time $t_{sleep}$ gets linearly increased by $t_{backoff}$ with every iteration (Line \ref{line:increment}) that contains an empty batch $b_s$. This process avoids any wastage of resources as there is no use of sending a request when the batch is empty. As soon as there are events in the batch, those are processed and the backoff interval is reset (Line~\ref{line:endbackThread}).

To receive events after they were processed from the $\omega_{UD}$ operator, we initialize another background thread (Line~\ref{line:startbackThreadRcv}--\ref{line:endbackThreadRcv}). Here also, we take advantage of command batching for a range query
(Line \ref{line:rangequery}). Much like database range queries,
the in-memory queue in \system provides the ability to define a start index and length to fetch items from the event queue. Using this command, we effectively reduce the number of requests to the event queue and fast-forwarding.
Similar to the send thread, the receive thread utilizes a linearly increased backoff time to avoid polluting an empty queue with requests (Line~\ref{line:incrementRcv}).

The combination of these mechanisms: parallel processing, manual flushing, and command batching, \system showed significant performance improvements in terms of throughput while still preserving low processing latency, which is evaluated in \Cref{sec:evaluation}.

%% file: sections/implementation.tex

%% file: sections/evaluation.tex
\section{Evaluation} \label{sec:evaluation}

In the evaluation  we intend to answer the following questions towards \system:

\begin{enumerate}
    \item How to provide dynamic operator updates using \system without runtime dependence? (\Cref{sec:performance-update})
    \item{What is the performance impact in terms of latency and throughput of implementing a user-defined operator using \system in comparison to a direct implementation in CEP systems?} (\Cref{sec:performance})
\end{enumerate}
We answer the above questions in a threefold evaluation.
In \Cref{sec:performance-update}, we evaluate the ability of \system to dynamically update an operator in comparison to the baseline of state-of-the-art CEP systems using throughput metrics.
In \cref{sec:performance}, we provide a performance comparison of \system with baseline CEP systems Flink and TCEP in terms of throughput (\Cref{sec:throughput-impact}) and latency (\Cref{sec:latency-impact}).

\begin{table}
\begin{center}
\small
\begin{tabular}{ p{4.5cm} p{3.5cm} }
\hline
Simulation time $t_s$ & 20 min \\
Warmup time $t_w$ & 60 s \\
Number of runs & 30 \\
Number of operators & 3 \\
Number of producers & \underline{1} - 3 \\
Number of brokers and consumers & 1 \\
Number of serverless operators & 1 \\
Back-off interval increment $t_{backoff}$ & 1 ns \\
Input event rate & \underline{1000}, 10,000 and 100,000 \\
\hline
CEP systems & \underline{Apache Flink}~\cite{flinkCarbone}, TCEP~\cite{Luthra2018} \\
Queries & Fraud detection
and \underline{forward} \\  
\hline
\end{tabular}
\setlength{\abovecaptionskip}{0pt}
\setlength{\belowcaptionskip}{-10pt}
\caption{Configuration parameters for the evaluation. \textnormal{\emph{Default or mostly used parameters are underlined.}}}
\label{tab:config-parameters}
\end{center}
\end{table}

\subsection{Evaluation Setup} \label{sec:evalSetup}
In the following, we describe the evaluation platform, the \system implementation, dataset, and queries used in the evaluation.

\textbf{Evaluation platform.} For evaluating \system, we utilize Docker version 19.03 running on a server with Intel(R) Xeon(R) CPU E5-2630 v2 @ 2.60GHz processors with Ubuntu 18.04 installed. The server has 128 GB RAM and 24 cores of Intel CPU.
We will use two different setup modes: (i) running direct implementation of $f_\omega$ in CEP system (baseline) and (ii) running the CEP system in combination with \system and a user-defined operator ($\omega_{UD}$).
For the former setup, we will execute the CEP system in a docker container which runs a given query. For the latter setup, the CEP system and CEPless will also be executed in two separate docker containers running the same query. The CEPless setup will be running the event queue, a user-defined operator $\omega_{UD}$, and the node manager on the same node, as proposed by our design. Furthermore, the docker containers are not restricted in their respective resource usage and therefore able to utilize the complete available node resources.

\textbf{Dataset.} We use a real-world dataset containing financial transactions \cite{transactionsDataset} to detect credit card fraud for evaluation. The anonymized dataset contains 284,807 rows of credit card transaction data. As a data producer, we used Apache Kafka running directly on the same computational resource.

\textbf{Queries.} The query used for the evaluation is shown in Figure \ref{benchmark-query}. In our evaluations, we will replace the \emph{Filter} ($\omega_\sigma$) by a user-defined operator realized in \system (figure at right). This operator performs a simple filter operation for filtering out transactions as fraud in the dataset. It will be evaluated using only the natively provided operator directly in the CEP system specification language and written as a $\omega_{UD}$ operator in our system.
Furthermore, to evaluate the performance impact of \system onto existing CEP systems, we use a forward query that forwards the event tuple or data stream directly to the consumer (cf. \Cref{sec:performance}).

We consider credit card terminals as producers (e.g. at supermarkets). Every card terminal produces an event as soon as a transaction process has started, i.e. the customer has presented the card at the terminal.  An emitted event contains the following tuples~\cite{transactionsDataset}: \textit{timestamp}, \textit{amount}, \textit{cardId}, \textit{terminalId}. Events sent by card terminals serve as input for a CEP query that detects fraudulent transactions. Terminals are connected directly through a wired connection to the network which enables stable connections to brokers.

\begin{table*}[]
\scriptsize
\begingroup
\setlength{\tabcolsep}{5.4pt} 
\begin{tabular}{l|l|l|l|l|l|l|l|l|l|l|l|l|}
\cline{2-13}
                                             & \multicolumn{4}{c|}{\textbf{1,000 events/s}}                                  & \multicolumn{4}{c|}{\textbf{10,000 events/s}}                                             & \multicolumn{4}{c|}{\textbf{100,000 events/s}}                                           \\ \hline
\multicolumn{1}{|l|}{\textbf{System}}        & \textit{mean} & \textit{min} & \textit{max} & \textit{quantiles (90, 95, 99)} & \textit{mean} & \textit{min} & \textit{max} & \textit{quantiles (90, 95, 99)} & \textit{mean} & \textit{min} & \textit{max} & \textit{quantiles (90, 95, 99)} \\ \hline
\multicolumn{1}{|l|}{\textbf{Flink}}         & 1047            & 108          & 1093         & 1001, 1002, 1002                & 10475           & 2180         & 19990        & 10010, 10011, 10020             & 100123           & 16770        & 200168       & 100098,100136,100797            \\ \hline
\multicolumn{1}{|l|}{\textbf{CEPless-Flink}} &  \textbf{1070}            &  \textbf{378}          &  \textbf{1190}         &  \textbf{1003, 1004, 1004}                &  \textbf{10477}            &  \textbf{1174}         &  \textbf{20007}        &  \textbf{10019, 10027, 12081}             &  \textbf{100353}           &  \textbf{18353}        &  \textbf{249689}       &  \textbf{100623,100815,101806}            \\ \hline
\multicolumn{1}{|l|}{\textbf{TCEP}} & 1021            & 396          & 2292         & 1001, 1002, 1002                & 10022            & 2660         & 45185        & 10010, 10015, 10110             & -           & -        & -       & -           \\ \hline
\multicolumn{1}{|l|}{\textbf{CEPless-TCEP}} & \textbf{1000}            & \textbf{214}          & \textbf{4927}         & \textbf{1001, 1002, 1004}                & \textbf{10002}            & \textbf{580}         & \textbf{32091}        & \textbf{10015, 10025, 10163}             & -           & -        & -       & -            \\ \hline
\end{tabular}
\endgroup
\setlength{\abovecaptionskip}{0pt}
\setlength{\belowcaptionskip}{0pt}
\caption{Throughput measurements: mean, min, max, and quantiles (90,95,99) for the forward operator.}
\label{tab:throughput}
\end{table*}
\begin{table*}[]
\scriptsize
\begin{tabular}{l|l|l|l|l|l|l|l|l|l|l|l|l|}
\cline{2-13}
                                              & \multicolumn{4}{c|}{\textbf{1,000 events/s}}                                     & \multicolumn{4}{c|}{\textbf{10,000 events/s}}                                    & \multicolumn{4}{c|}{\textbf{100,000 events/s}}                                  \\ \hline
\multicolumn{1}{|l|}{\textbf{System}}        & \textit{mean} & \textit{min} & \textit{max} & \textit{quantiles (90, 95, 99)} & \textit{mean} & \textit{min} & \textit{max} & \textit{quantiles (90, 95, 99)} & \textit{mean} & \textit{min} & \textit{max} & \textit{quantiles (90, 95, 99)} \\ \hline
\multicolumn{1}{|l|}{\textbf{Flink}}         & 0.78            & 1.0          & 31.46        & 0.98, 1.06, 1.55                & 1.71            & 1.0          & 341.61       & 2.09, 2.22, 3.11                & 1.74            & 1.0          & 87.91        & 2.29, 2.45, 2.88                \\ \hline
\multicolumn{1}{|l|}{\textbf{CEPless-Flink}} & \textbf{3.19}            &  \textbf{1.0}          &  \textbf{26.19}        &  \textbf{4.10, 4.35, 4.9}                &  \textbf{3.63}            &   \textbf{1.15}         &  \textbf{503.50}       &  \textbf{4.21, 4.48, 5.15}                &  \textbf{4.61}           &  \textbf{1.82}         &  \textbf{956.31}       &  \textbf{5.71, 6.11, 7.80}             \\ \hline
\multicolumn{1}{|l|}{\textbf{TCEP}} & 1.14            & 1.0          & 27.89         & 1.45, 1.57, 1.82                & 4.21            & 1.0         & 496.63        & 2.54, 2.74, 24.32             & -           & -        & -       & -           \\ \hline
\multicolumn{1}{|l|}{\textbf{CEPless-TCEP}} & \textbf{4.94}            & \textbf{0.72}          & \textbf{21.27}         & \textbf{6.10, 6.37, 6.86}                & \textbf{5.21}            & \textbf{1.97}         & \textbf{488.84}        & \textbf{6.25, 6.75, 10.38}             & -           & -        & -       & -            \\ \hline
\end{tabular}
\setlength{\abovecaptionskip}{0pt}
\setlength{\belowcaptionskip}{-15pt}
\caption{Latency measurements: mean, min, max, and quantiles (90,95,99) for the forward operator (in ms).}
\label{tab:latency}
\end{table*}

A summary of the configuration parameters we used in the evaluation can be found in Table \ref{tab:config-parameters}.
\emph{Backoff time} describes the interval in which the aforementioned batches are written to the network layer.
In the next section, we go into the performance evaluation of \system.

\subsection{Evaluation of dynamic operator updates}\label{sec:performance-update}
To understand how CEP systems benefit from \system operator updates, we show evaluations regarding the explicit time when updating an operator that is currently in execution. For this, we utilize two metrics: i) downtime, ii) update time and (iii) throughput. \emph{Downtime} represents the amount of time where no events were received at the consumer side, while \emph{update time} gives the amount of time it took to update the operator to the new version. \emph{Throughput} is defined as the number of output events received at the consumer for the input events ingested by the producer.  Such operator updates are necessary for applications like fraud detection, earlier motivated in \Cref{subsec:example}. Small down and update times are considered to be good. We evaluated \system in comparison to Apache Flink by performing an operator update using our presented mechanism for the former and by deploying a query with the new operator for the latter. The business logic of the operator gets updated to detect \emph{more} fraud patterns hence resulting in a higher throughput of events after the update. Ideally, the throughput should remain at a constant rate to not have performance peaks in resource usage of the given server. Both systems share the same business logic inside the operator before the update and get updated with the same business logic.

\begin{figure}
    \includegraphics[width=0.8\linewidth] {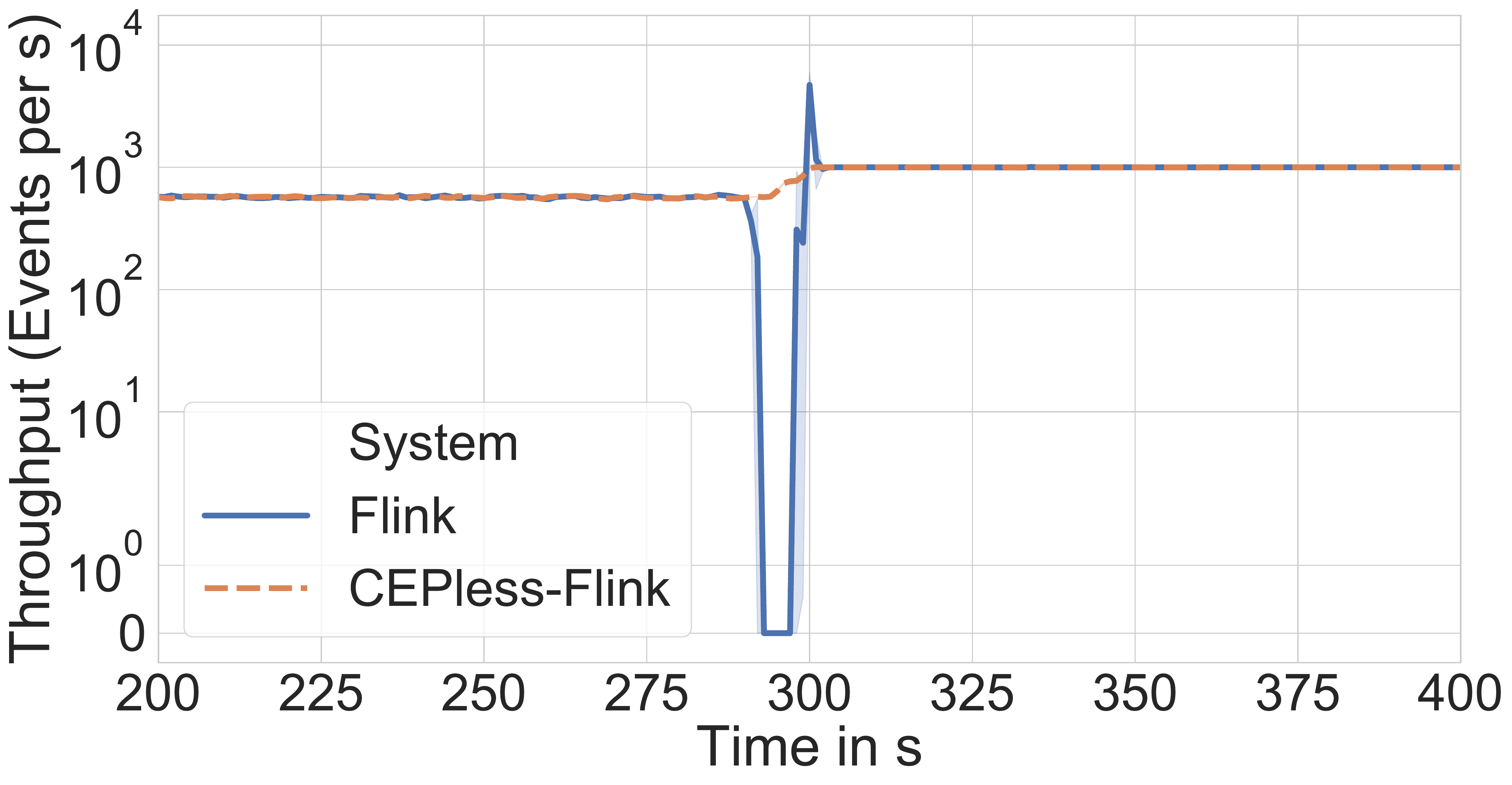}
     \setlength{\abovecaptionskip}{0pt}
    \setlength{\belowcaptionskip}{-5pt}
   \caption{Apache Flink experiences a downtime while \system steadily updates the operator while providing optimal throughput.
   }
    \label{fig:operator-update}
\end{figure}
The evaluated throughput of the experiment is shown in Figure \ref{fig:operator-update}. We repeated the experiment 30 times and plot the 95\% confidence interval (shown as shadow in the line plot) of the observations. The operator update was issued at t = 290 in both systems. While the update is proceeding in Flink a downtime is observed, because of the deployment of the new query. This downtime is introduced by the distribution and execution of the used JAR-file for the query. The mean downtime of the query was approximately \textbf{8.6 s}, which is equivalent to the update time, and resulted in a throughput decline and spike at approximately t = 298. This spike is caused by the events that were not processed while the query was in an update state. As soon as the new query is executing again, it begins processing from the last saved checkpoint at the data generator.
In comparison, CEPless showed no downtime at all, only a mean update time of \textbf{238~ms} without a throughput decline or spike. In fact, the \system achieves a throughput of the new query very steadily at t = 300.

We achieved this relatively small update time with the mechanism of updating the  operator containers instead of complete queries. While this update occurs, the state of the query currently in execution is kept in the presented message queues.

\subsection{Performance Evaluation} \label{sec:performance}

To understand how the \system system influences a CEP system we analyze both implementations (Apache Flink and TCEP) in terms of \emph{end-to-end latency} and \emph{throughput}. These two are the most important metrics for a CEP system as well as the financial fraud detection example. We evaluate a forward query introduced in \Cref{sec:evalSetup} to observe whether \system achieves optimal throughput and latency while forwarding the data streams.
 We evaluate the following three system configurations:

\begin{enumerate}
\item Running only the native CEP system implementation without our extension (baseline).
\item Running the CEP system with a user-defined operator, Redis \cite{redis} as an in-memory queue with and without batching.
\end{enumerate}

The impact of throughput and latency are evaluated on Flink and TCEP that showed similar observations as detailed in the following. This shows the universal applicability and behavior of different CEP systems.

\subsubsection{Impact on throughput}\label{sec:throughput-impact}
 An \emph{optimal} throughput is achieved when the output events matches the input event rate. We collected throughput measurements for the above three configurations for 20 minutes and other parameters as presented in \Cref{tab:config-parameters}. We repeated the experiment 30 times to provide 90, 95, 99\% confidence interval of throughput for the different input rates, respectively. 
 In \Cref{tab:throughput}, we present the mean throughput measurements
 for the baseline Flink and our extension \system with Flink ingested with an input rate of 1,000, 10,000, and 100,000 events per second, respectively from left to right. In comparison, our extension of Flink with \system achieves matching optimal throughput per time unit for the given input rate.
 Clearly, \system-Flink matches the baseline in all the results.
 We observed similar results for our extension on TCEP for throughput. However, TCEP baseline was not able to perform well under a higher scale (100K events per second) because of the absence of back pressure and flow control mechanisms in the system. Hence we do not report the results for the 100K event rate. For 1K and 10K events, \system with TCEP performs equally well and attains optimal throughput. Henceforth, our extension \emph{does not} introduce any overhead in terms of throughput and can easily deal with a high scale of ingested events as seen in the evaluations.

 To achieve matching throughput for high event rates 100K events we set the input batch size to 10,000. As described in Section \ref{sec:in-memory-queue}, this implicitly increases the size of a range query in Redis leading to lower queue processing overhead due to fewer requests and internal network round-trips. We further elaborate on the batch size in the last subsection.

\Cref{tab:throughput} in some cases also shows higher throughput results for \system-Flink than Apache Flink (baseline), which is an interesting observation. We analyzed this behavior and find out that since we process events in batches due to the batching mechanism introduced in \Cref{alg:flush}, the events are queued up in the batch instead of getting processed directly. Moreover, Flink implements flow control and backpressure mechanisms, in particular, credit based flow control~\cite{flinkCarbone} to deal with such situations.
The throughput for \system-Flink fluctuates much from the optimum as the system processes events that were previously received but backpressured by the Flink engine. Therefore, when looking at the evaluated values the throughput per second seem higher at times, but at minimal latency cost.

\subsubsection{Impact on latency} \label{sec:latency-impact}
The \emph{end to end latency} is defined as the time taken to retrieve an event from the producer until it reaches the consumer from the CEP engine. Similar to the throughput evaluations, we collected latency measurements for different input event rates.
In \Cref{tab:latency}, we present the latency measurement observed for the Flink baseline and our extension. Our \system system atop Flink attains good performance while introducing only a minimal overhead in terms of latency. 
We observe only a mean overhead in latency of \textbf{1.92 ms} for 10,000 events per second and  \textbf{2.87 ms} for 100,000 events per second (calculated as \emph{mean latency of \system -- mean latency of Flink}). 
Also for the TCEP baseline, similar to Apache Flink, our extension here as well induce a minimal overhead of \textbf{1~ms} for 10K events.  The overhead is minimal because we have parameterized the batch sizes for a suitable configuration.
The overhead comprises of two factors: i) the translation in the universal message format of every event and ii) the round-trip-time (RTT) of the message queue to the CEP system in use.

%% file: sections/related-work.tex

\section{Related Work} \label{sec:relatedwork}

In the following, we present previous work in terms of flexible operator deployment and unifying CEP systems and serverless frameworks.

\textbf{\textit{CEP Systems and Programming Models}}. Apache Beam \cite{beam} developed by Google provides a unified programming model as an abstraction layer above multiple different CEP systems including Apache Flink \cite{flinkCarbone} and Storm~\cite{apacheHeron}. TCEP~\cite{Luthra2018} and ProgCEP~\cite{Luthra2019} proposes a programming model for operator placement in CEP systems.
These works confirms the need to have a unified programming model for deploying queries on different CEP systems.
However, with operators being bound to the respective execution environments -- as proposed in Beam -- CEP systems fail to fulfill the flexibility of complex applications that require different systems to interact together.
Recently, Bartnik et al.~\cite{runtimeflink/datenbank/Bartnik2019} proposed an extension of Flink with runtime updates of operators, however, their proposal is Flink's runtime dependent and not applicable to other CEP runtime systems. Some classical systems like Amit~\cite{amit/vldb/asaf2004} and Apama~\cite{apama} allow dynamic updates of CEP operators, however, with a strict dependency on the runtime. Many CEP languages have been developed in the past years such as CQL~\cite{cql/vlbd/arasu2006},
SASE~\cite{sase/sigmod/eugene2006}, and TESLA~\cite{tesla/debs/cugola2010}, besides the domain-specific ones proposed in the above programming models.
However, as per our knowledge, none of the above programming models provide a powerful abstraction on multiple runtime environments as we do.

\textbf{\textit{Serverless Frameworks}}. In recent years, many commercial serverless platforms evolved e.g., AWS Lambda~\cite{lambda}, Google Cloud Functions~\cite{googleCloudFunctions}, Azure Functions~\cite{azureFunctions}, and IBM OpenWhisk~\cite{whisk}. Most of these providers often support the streaming of input data towards serverless functions, making it possible to also execute continuous data flows. However, often those services are limited to the provider-specific streaming solutions for example Kinesis \cite{kinesis} by Amazon. An open-source alternative to AWS Lambda is Kubeless~\cite{kubeless}, which provides multiple different runtime environments supporting different languages, but also the ability to provide a custom runtime using a custom image. Programming models for serverless~\cite{joyner:arxiv:2020:ripple, McGrath2017} on top of AWS  and Microsoft Azure, respectively have been proposed as well. However, an integrated solution, with CEP systems, \eg Flink, is still missing which we see as a gap in current serverless execution paradigms. Our system could be used as another serverless service by cloud providers to provide real serverless operators in combination with existing streaming systems.

%% file: sections/conclusion.tex
\section{Conclusion}~\label{sec:conclusion}
In this work, we proposed \system, a CEP system based on \emph{serverless computing}, which provides flexibility in developing \emph{new} user-defined operators in any programming language and updating them at runtime. This is highly beneficial for applications that require dynamic changes such as fraud detection in financial context and Internet of Things applications. To this end, we
(i) ease development, (ii) provide dynamic operator updates and (iii) improve the flexibility of operators in CEP system while preserving the statefulness and meeting the performance requirements in terms of throughput (up to 100K events per second) and latency (\textasciitilde{1.9}~ms).
This work essentially takes an important step towards \emph{serverless computing} for CEP systems.
Furthermore, it can be highly beneficial to develop user-defined operator placement algorithms that are independent of the execution environment. Besides, distributed in-memory queues such as Anna~\cite{Wu2018AnnaQueue} can be used to achieve a higher scale in stateful processing for instance for bigger window sizes.
Finally, \system provides a foundation for developing advanced serverless features like just-in-time billing and auto-scaling on cloud platforms. For example, the programming interface can provide the information required for billing operators and container technology is at the core of \system can be easily used for auto-scaling. Sophisticated scaling strategies for CEP~\cite{Gulisano2012} can also be integrated as a runtime in our execution layer. 